\documentclass[prd,aps,preprintnumbers,showpacs]{revtex4}
\usepackage{epsfig}
\usepackage{graphicx}
\usepackage{bm}

\newcommand{\be}{\begin{equation}}

\newcommand{\ee}{\end{equation}}
\newcommand{\bea}{\begin{eqnarray}}
\newcommand{\eea}{\end{eqnarray}}

\def\starteq{
\begin{eqnarray}}
\def\fineq{\end{eqnarray}}

\def\a{\alpha}
\def\b{\beta}
\def\d{\delta}

      \def\G{\Gamma}

\def\l{\lambda}

\def\p{\pi}      
   
\def\r{\rho}
\def\s{\sigma}

\def\rarr{\rightarrow}

\begin{document}
\preprint{BARI-TH 496/04}
\preprint{DSF-2004/25 (Napoli)}

\title{Final state interactions for $B\to VV$ charmless decays}

\author{\textbf{Massimo Ladisa, Vincenzo Laporta, Giuseppe Nardulli}}
\affiliation{Dipartimento di Fisica dell'Universit{\`a} di Bari, Italy\\
Istituto Nazionale di Fisica Nucleare, Sezione di Bari, Italy}
\author{\textbf{Pietro Santorelli}}
\affiliation{Dipartimento di Scienze Fisiche, Universit{\`a} di
Napoli "Federico II", Italy\\
Istituto Nazionale di Fisica Nucleare, Sezione di Napoli, Italy}

\begin{abstract}
We estimate final state interactions in the
$B$-meson decays into two light vector mesons by the Regge model.
We consider Pomeron exchange and charmed Regge trajectories that
can relate intermediate charmed particles to the final state. The
Regge poles have various helicity-flip residues, which allows a
change from the longitudinal to transverse  polarization. In this
way a significant reduction of the longitudinal polarization
fraction can be produced. In the factorization approximation we
find agreement with recent data from the BaBar and Belle
collaborations in the $B\to K^*\phi$ decay channel, as a result of
an appropriate choice of semileptonic form factors and Regge
exchanges. On the other hand, data for the $K^*\r$ decay channels
appear more elusive. The soft effects discussed in the present
paper are based on a model of Regge trajectories that is shown to
reproduce correctly in the non-charmed case the Regge
phenomenology of light mesons.
\end{abstract}

\maketitle

\section{Introduction \label{sec:-1}}
Recent data from the Babar and Belle collaborations
\cite{Aubert:2003mm,Aubert:2003xc,belle:2003jf,Zhang:2003up} on
the $B$ decays into two light vector mesons $B\to V\,V$ have
produced considerable theoretical interest, see e.g. the recent
papers
\cite{Kagan:2004uw,Grossman:2003qi,Colangelo:2004rd,Kagan:2004ia}.
Data for the Branching Ratios (BR) and  the polarization fractions
for a few decay channels are reported in Table~\ref{table0}.
\begin{table}[h]
\begin{center}
\begin{tabular}{|c|c|c|c|c|}
\hline Decay mode &BR & $\Gamma_L/\Gamma$&$\Gamma_\perp/\Gamma$
&Ref.\\
\hline
$B^+\to\rho^0 K^{*+}$&$(10.6^{+3.0}_{-2.6}\pm2.4)\times10^{-6}$\,
&$0.96^{+0.04}_{-0.15}\pm0.04$&
&\cite{Aubert:2003mm},\cite{Aubert:2003xc}\\
\hline$B^+\to\phi K^{*+}$&$(9.5\pm1.7)\times10^{-6}$&$0.46\pm0.12\pm0.03$&
&\cite{Aubert:2003mm},\cite{belle:2003jf}\\
\hline $B^0\to\phi
K^{*0}$&$(10.7\pm1.2)\times10^{-6}$&$0.58\pm0.06$&$0.41\pm0.10\pm0.02$
&\cite{Aubert:2003mm},\cite{belle:2003jf}\\
\hline $B^+\to\rho^0
\rho^{+}$&$(26.2\pm6.2)\times10^{-6}$&$0.96\pm0.07$&
&\cite{Aubert:2003mm},\cite{Aubert:2003xc},\cite{Zhang:2003up}\\
\hline $B^0\to\rho^+
\rho^-$&$(25^{+7+5}_{-6-6})\times10^{-6}$&$0.98^{+0.02}_{-0.08}\pm0.03$&
&\cite{Aubert:2003mm},\cite{Aubert:2003xc}\\
\hline
\end{tabular}
\end{center}\caption{
Survey of experimental results for $B$ decays into two light
vector mesons. Data  of Refs.\cite{Aubert:2003mm} and
\cite{Aubert:2003xc} are  from the BaBar
collaboration. Data from Refs. \cite{belle:2003jf} and \cite{Zhang:2003up}
are from the Belle Collaboration. There is also an
upper bound ${\cal B}(B^0 \to \rho^0
\rho^0)\le 2.1 \times 10^{-6}$ from BaBar
\cite{Aubert:2003mm}.
\label{table0}}
\end{table}

 In these decay modes the two vector mesons have the same helicity; therefore three
 different polarization states are possible,
 one longitudinal ($L$) and two transverse, corresponding to helicities $\l=0$ and
 $\l=\pm 1$. We define the corresponding amplitudes as $A_{0,\pm}$. It is easy to
 show \cite{Kagan:2004uw} that in the large $m_b$ limit longitudinal
 polarization must be enhanced by a factor of $m_b$ in the amplitude.
 This can be seen at the quark level, the reason being that transverse
 polarizations are generated by helicity flip and this implies an extra
 factor $m_V/m_b$. Alternatively, at the hadronic level,
 by making use of naive factorization, one traces back the enhancement of
 the longitudinal polarization to the presence of an extra factor
 $m_B/m_V+{\cal O}((m_V/m_B)^2)$ in the longitudinal polarization vectors.
 The enhancement in the longitudinal amplitude
 is of one power of $m_b$. In fact the term with
two powers is multiplied by the difference
$A_1-A_2$ where  $A_1$ and $A_2$ are the usual axial form factors \cite{Bauer:1986bm}
for the transition
 $B\to V$ computed at $q^2=m_{V}^2$. This difference vanishes in the
 high $m_b$ limit, see e.g. \cite{Charles:1998dr}.

Decay modes $B^0 \to \rho^+ \rho^-$
and  $B^+ \to \rho^0 \rho^+, \rho^0 K^{*+}$ confirm this prediction
as the final states are
dominated by the longitudinal polarization. On the other hand,
for the observed $B \to \phi K^*$ transitions, the longitudinal
amplitude width gives about $~50\%$ of the rate. This is a major puzzling
 feature because, as argued above, in the naive factorization,
one expects also for the $\phi K^*$ channels
\be
\frac{\Gamma_L}{\Gamma}=\frac{|A_0|^2}{|A_0|^2+|A_+|^2+|A_-|^2}=
1-{\cal O}\left(\frac{1}{m^2_b}\right)\ .
\ee
Another surprising aspect concerns the transverse polarization defined by
\be
\Gamma_\perp\propto |A_\perp|^2\, ,
\ee
where $A_{\perp,\parallel}=(A_+\mp A_-)/\sqrt 2$. Moreover, one expects \cite{Kagan:2004uw}
\be
 \frac{\Gamma_\perp}{\Gamma_\parallel}=
 1+{\cal O}\left(\frac{1}{m_b}\right)
 \label{naive}\, .
 \ee
 This
 expectation is based on naive factorization and large energy  relations
 \cite{Charles:1998dr}. The former assumption implies that
 in the large $m_b$ limit  $A_-$ is proportional to $(A_1-V)/(A_1+V)+{\cal O}(m_V/m_B)$ where
 $A_1$ and $V$ are the usual form factors \cite{Bauer:1986bm} for the transition
 $B\to K^*$ computed at $q^2=m_{\phi}^2$. The latter implies that
$(A_1-V)/(A_1+V)={\cal O}(1/m_b)$, a prediction confirmed
by different numerical computations
based on lattice QCD or QCD sum rules. As a consequence $A_-/A_+={\cal O}(1/m_b)$
and (\ref{naive}) should hold. A result from Belle is reported in
Table \ref{table0}, which is at odds with (\ref{naive}); a smaller
result from BaBar \cite{Gritsan:2004new}
\be
\frac{\Gamma_\perp}{\Gamma}=0.27\pm 0.07\pm 0.02
\label{babarr}
\ee
has been also presented, which might be compatible with (\ref{naive}).

All previous considerations do not take into account final state
interactions (FSI) that, although power suppressed, might nevertheless
produce sizeable effects. In \cite{Colangelo:2004rd}
a parametrization of these effects is considered. It assumes a major role of
the so-called charming penguin diagrams. They are
Feynman diagrams with  intermediate $D\bar D$ states
\cite{Colangelo:1989gi,Ciuchini:1997hb,Isola:2001ar,Isola:2001bn,Isola:2003fh}.
The suppression due to their non-factorizable status might be
compensated by the Cabibbo-Kobayashi-Maskawa (CKM) enhancement.
The computational scheme used in \cite{Colangelo:2004rd}
and in previous computations for $B$ decays into two light particles
\cite{Isola:2001ar,Isola:2001bn,Isola:2003fh} is chiral perturbation
theory for light and heavy mesons (for a review see \cite{Casalbuoni:1997pg}).
Since  momenta of the final particles are hard, the application of the method
involves large extrapolations and the method is questionable
(see e.g. \cite{Kagan:2004ia}). A possible answer consists of  introducing
some correction. In \cite{Isola:2001ar,Isola:2001bn,Isola:2003fh} such correction
was modelled by a form factor evaluated by a constituent quark model. In
\cite{Colangelo:2004rd} a phenomenological parameter $r$ is introduced to weight
the charming penguin contribution. Both approaches result in a suppression
of charming penguin contributions, but uncertainties are  huge.

The reason for the suppression can be traced back to the Regge theory.
In fact, given the rather large energy involved (${\sqrt s}=m_B$),
the Regge approach should be
a good approximation for the  final state interactions and provide
 a computational scheme of the rescattering effects,
 be they elastic or inelastic. The former are described by Pomeron exchange
  \cite{Donoghue:1996hz}, \cite{Nardulli:1997ht}.
 Among the latter, one may still assume dominance
 of the charming penguin amplitudes, due to the CKM enhancement, but
 they should be evaluated {\it via} a reggeized amplitude.
In this context the suppression of the charming penguin
arises because the Regge charmed
trajectories have a negative intercept $\alpha(0)$ and therefore a suppression factor
$(s/s_0)^{\alpha(0)}$ ($s_0\simeq 1$ GeV$^2$, a threshold). The advantage of the
Regge approach is to evaluate the intermediate charmed states contributions
not by Feynman diagrams, but by unitarity diagrams and the Watson's theorem \cite{Watson:1952ji}.
In this way the remarks connected to the extrapolation of chiral theory to
hard momenta are by-passed.

The aim of this paper is to give an evaluation of the final
 state interactions for $B\to VV$ using
the Regge model. Clearly also the numerical values of the bare (no
FSI) amplitudes are important. We compute them in the
factorization approximation and therefore we need a set of
semileptonic form factors as an input. We use a determination
based on Light Cone QCD sum rules \cite{Ball:2003rd} that
encompasses both the $B\to K^*$ and the  $B\to\r$ transitions. For
$B\to K^*$ we also use results obtained by QCD sum rules
\cite{Colangelo:1996jv}. These results are summarized in section
\ref{sec:0}.
 In section \ref{sec:1} we discuss
rescattering effects as parametrized by the Regge model.  We
include in the parametrization the Pomeron and charmed Regge
trajectories and we give estimates of the Regge residues. In
section \ref{sec:2} we present our numerical results and discuss
them. The basic result we find is that, due to the presence of
helicity-flip Regge residues, longitudinal polarization fractions
are in general reduced, except for  the $B\to\r\r$ decay modes.
Differences between the channels $B\to K^*\phi$ and $B\to K^*\r$
might depend on the choice of the form factors, therefore we
discuss differences induced by this choice as well. We argue that
for the $B\to K^*\phi$ decay mode an interplay between form
factors and Regge parameters can produce theoretical results
compatible with the data with no need to introduce new physics
effects. For the $B\to K^*\r$ the results are more elusive and
more work is needed especially for the determination of the
semileptonic form factors. Finally in the Appendix we give an
outline of the model used to compute the parameters of the charmed
Regge trajectories.
\section{Bare amplitudes \label{sec:0}}
In this section we introduce bare amplitudes, i.e.
matrix elements of the weak hamiltonian with no final state interaction effects.
We consider two sets of weak bare amplitudes. The first set includes
the bare amplitudes for the $B$ decays into two light vector mesons.
 We consider the channels $B\to \r\r$, $B\to K^*\r$, $B\to K^{*}
 \phi$. Since this calculation is straightforward
 we limit our presentation to the numerical results.

For the relevant Wilson coefficients we use
  $a_1=1.05$, $a_2=0.053$,  and
 $(a_3,a_4,a_5,a_7,a_9,a_{10})$ $=(48,-439-77i,-45,+0.5-1.3i,-94-1.3i,-14-0.4i)$
  $\times 10^{-4}$ \cite{Ali:1998eb}. We use  two different choices of form factors
 for the weak transition $B\to V$. The first choice is based on Light Cone QCD sum rules
 \cite{Ball:2003rd}. The corresponding results for the bare amplitudes for $B\to K^*\r$
 and $B\to \r\r$
 are reported in  Table \ref{table1bis}. We also consider the QCD Sum Rules results of Ref.
 \cite{Colangelo:1996jv}. Since in this paper only the $b\to s$ transitions
 are considered, these results
 can be employed only for $B\to K^*\phi$ decays. The numerical results obtained
 by the two sets of form factors for the  $B\to K^*\phi$ decay channel are reported in Table
 \ref{table1ter}.
\begin{table}[ht]
\begin{center}
\begin{tabular}{||c|c|c||c|c|c||}
\hline
Process & $\lambda$& $A_{b,\lambda}$
&Process& $\lambda$&$A_{b,\lambda}$\\
\hline
$B^+\to K^{*0}\rho^+$&0&$-0.62+3.14\,i$ &
$B^+\to \r^{0}\rho^+$&0&$+4.21-2.60\,i$
\\
\hline
$B^+\to K^{*0}\rho^+$&+&$+0.17-0.86\,i$&
$B^+\to \r^{0}\rho^+$&+&$-1.01+0.62\,i$\\
\hline
$B^+\to K^{*0}\rho^+$&-&$-0.02\,i$ &
$B^+\to \r^{0}\rho^+$&-&$-0.01+0.01\,i$\\
\hline
$B^+\to K^{*+}\rho^0$&0&$+0.54+2.64\,i$&
$B^0\to \r^{+}\rho^-$&0&$+5.60-3.93\,i$\\
\hline
$B^+\to K^{*+}\rho^0$ &$+$&$-0.15-0.70\,i$
& $B^0\to \r^{+}\rho^-$&+&$-1.34+0.94\,i$\\
\hline
$B^+\to K^{*+}\rho^0$&-&$-0.01\,i$&
$B^0\to \r^{+}\rho^-$&-&$-0.02+0.01\,i$\\
\hline
\end{tabular}
\end{center}\caption{
Bare helicity amplitudes for
$B\to K^*\r$ and $B\to \r\r$ with form factors from Ref. \cite{Ball:2003rd}.
Results in $10^{-8}$ GeV.
\label{table1bis}}
\end{table}

\begin{table}[ht]
\begin{center}
\begin{tabular}{||c|c|c||c||}
\hline Process & $\lambda$ &
$A_{b,\lambda}$(Ref. \cite{Ball:2003rd}) &
$A_{b,\lambda}$(Ref. \cite{Colangelo:1996jv})\\
\hline
$B\to K^*\phi$  & $0$ & $-0.83+3.85\,i$ &  $-0.48+2.24\,i$\\
\hline
$B\to K^*\phi$  & $+$ & $+0.27-1.25\,i$ &  $+0.28-1.30\,i$\\
\hline
$B\to K^*\phi$  & $-$ & $-0.03\,i$      &  $+0.01-0.06\,i$\\
\hline
\end{tabular}
\end{center}\caption{
Bare helicity amplitudes for $B\to K^*\phi$ with two different choices
of form factors \cite{Ball:2003rd} and \cite{Colangelo:1996jv}.
Results in $10^{-8}$ GeV.
\label{table1ter}}
\end{table}
We note the numerical relevance of
$A_0$ and  the suppression of $A_-$ in comparison with $A_+$. We
 also note numerical differences
in Table \ref{table1ter} arising from the different choice of form factors.

The second set of amplitudes includes the bare decays of $B^+$ into two charmed mesons,
$D_s^{(*)+}\, {\bar D}^{(*)0}$. For the $B\to\r\r$ decay channel we consider
$B$ decay into two charmed non strange mesons. The matrix elements
of this second set are computed in the factorization
approximation, using the Heavy Quark Effective Theory at leading
order, with the following  Isgur-Wise function
$\xi(\omega)=4/(1+\omega)^2$. For the CKM matrix elements we use
PDG data \cite{Eidelman:2004wy}.

Numerical results for the  bare helicity amplitudes
$A_{b,\lambda}$ for the channels $B^+\to D_s^{(*)+} {\bar
D}^{(*)0}$
 are reported in Table  \ref{table1} .

\begin{table}[!!!!h]
\begin{center}
\begin{tabular}{||c|c|c||}
\hline
Process&$\lambda$&$A^{(k)}_{b,\lambda}$\
\\
\hline
$B^+\to D^+_s\bar D^{0}$&0&$+1.32\,i$
\\
\hline
$B^+\to D^+_s\bar D^{*0}$&0 & $-1.15\,i$\\
\hline
$B^+\to D_s^{*+}\bar D^{0}$&0 &$-1.15\,i$\\
\hline
$B^+\to D_s^{*+}\bar D^{*0}$&0 &$+1.32\,i$\\
\hline $B^+\to D_s^{*+}\bar D^{*0}$&$+$&$-1.02\,i$\\
\hline $B^+\to D_s^{*+}\bar D^{*0}$ &$-$
& $-0.42\,i$ \\
\hline
\end{tabular}
\end{center}\caption{
Bare helicity amplitudes for $B$ decays into charmed mesons. The four channels
$B^+\to D_s^{+}\bar D^{0}$, $B^+\to D_s^{*+}\bar D^{0}$, $B^+\to D_s^{*+}\bar D^{0}$, and
  $B^+\to D_s^{*+}\bar D^{*0}$ are
identified respectively by the index $k=1,2,3,4$. Units are $10^{-6}$ GeV.
\label{table1}}
\end{table}

One can immediately notice the CKM enhancement (typically by a
factor $10^2$) of the amplitudes in Table \ref{table1} in
comparison with those of  Table \ref{table1bis}.  For $B$ decay
into two non-strange charmed mesons the enhancement is smaller.

\section{Final state interactions and Regge behavior \label{sec:1}}
 Corrections to the bare amplitudes due to final state interactions are taken into
account by means of the Watson's theorem \cite{Watson:1952ji}:
\be
A = \sqrt{S}A_{b}
\label{amp}
\ee
where $S$ is the
$S$-matrix, $A_{b}$ and $A$ are the bare and the full amplitudes.

The two-body $S$-matrix elements are given by
\starteq
S^{(I)}_{ij} = \d_{ij} + 2i\sqrt{\r_{i}\r_{j}}\
T^{(I)}_{ij}(s)~~,
\label{s-matrix}
\fineq
\par\noindent
where $i,j$ run over all the channels involved in the final state interactions.

The $J=0$, isospin $I$ amplitude $
T^{(I)}_{ij}(s)$ is obtained by projecting the $J=0$ angular
momentum out of the amplitude $T^{(I)}_{ij}(s,t)$:
\starteq
T^{(I)}_{ij}(s) =  {1 \over 16\p}{s \over \sqrt{\ell_{i}\ell_{j}}}
\int^{t_{-}}_{t_{+}}dt \,T^{(I)}_{ij}(s,t) ~~.
\label{S-wave}
\fineq
The momenta $\r_j\,,\ell_j$ are defined below.
For the channel $B\to K^*\phi$ we only have the $I=1/2$ transition amplitudes, for
$B\to K^*\r$ both $I=1/2$ and $I=3/2$ are involved. For $B\to \r\r$ we can have $I=0$
and $I=2$.

As discussed in the Introduction, given the rather large value of
$s=m^2_B$, a Regge approximation for the transition amplitude
seems adequate. We will therefore include first of all the Pomeron
term, which contributes to the elastic channels. For the inelastic
channels we include only channels whose bare amplitudes are
prominent. For $B\to K^*\phi$ and $B\to K^*\r$ they should be  the
amplitudes for the transition $B\to D^{(*)} D_s^{(*)}$ since these
channels are Cabibbo enhanced, see table \ref{table1}. For the
$I=0$ amplitude of the decay $B\to\r^+\r^-$ there is no Cabibbo
enhancement; most probably other trajectories should be
considered, but we consider here again charmed trajectories mainly
 for the sake of comparison.   The
transition from a state with two charmed mesons to a state with
two light particles can only occur {\it via} charmed Regge poles.
Therefore in the present approximation we will include, besides
the Pomeron, only the charmed Regge trajectories.

\subsection{Pomeron contribution}
To begin with we  consider the Pomeron contribution. We write
\be
S={\cal P}
\label{P0}
\ee
neglecting for the time being
non leading Regge trajectories and inelastic terms. We have
\be
{\cal P}=1+2iT^{{\cal
P}}(s)\,,\hskip1cm T^{{\cal P}}(s)=\frac{1}{16\p s
}\int_{-s+4m^2_V}^{0}T^{{\cal P}}(s,t) dt\, .
\label{p1}
\ee
The following parametrization can be used
\cite{Nardulli:1997ht}, \cite{Donnachie:1992ny}:
\starteq
T^{{\cal P}}(s,t)=~-~ \b^{P}g(t) \left({s \over s_{0}}
\right)^{\a_{P(t)}} e^{ -i{\p \over 2}\a_{P}(t)}~~,
\label{P}
\fineq
with $s_{0} = 1\,\rm GeV^{2}$ and
\starteq
\a_{P}(t) = 1.08
+ 0.25t   \qquad \qquad (t {\rm \ in \ GeV^{2}})~~,
\label{alphaP}
\fineq
as given by fits to hadron-hadron scattering total cross
sections. The product $\b^{P}\cdot
g(t) = \b^{\cal P}(t)$ represents the Pomeron residue; for the
$t$-dependence we assume \cite{Nardulli:1997ht}, \cite{Donnachie:1992ny}
\starteq
g(t) = {1 \over (1-
t/m_{\r}^{2})^{2}} \simeq e^{2.8t}~~.
\label{gt}
\fineq
To get $\b^{\cal P}$ we use factorization and
the additive
quark counting rule.
For the $B\to K^*\rho$ channel residue factorization gives
 \be
 \b^{\cal P}_{K^*\rho}=\b^{\cal P}_\r\b^{\cal
P}_{K^*}\ .
\label{fa}
\ee
The two residues appearing in (\ref{fa}) can be computed by the additive
quark counting rule. This gives \cite{Nardulli:1997ht}
\bea
\beta^{\cal P}_{\r}&\sim& \beta^{\cal P}_{\p}=2\b^{\cal P}_{u}\, ,\nonumber\\
\beta^{\cal P}_{K^*}&\sim& \beta^{\cal P}_{K}=\b^{\cal P}_{u}+
\b^{\cal P}_{s}\, .
\eea
The basis of the additive quark counting rule is given
 by pion-proton and proton-proton high energy scattering data.
From these data  \cite{Nardulli:1997ht},\cite{Zheng:1995mg}:
$\b^{\cal P}_{\p}\sim 2/3 \b^{\cal P}_{p}\sim 5.1$.
From $Kp$ high energy scattering data, one finds
$\b^{\cal P}_{s}\sim 2/3\b^{\cal P}_{u}$. Therefore
\be
\b^{\cal P}_{K^*\rho}\approx 22\ .
\ee
For the $B\to K^*\phi$ channel we have, instead of
 Eq. (\ref{fa}), $ \b^{\cal P}_{K^*\phi}=
\b^{\cal P}_\phi\b^{\cal
P}_{K^*}\ , \label{fa1}$ and, numerically,
$ \b^{\cal P}_{K^*\phi}\approx 14\ .
 \label{fa10}$
Finally for the $B\to \rho\rho$ channel we get
$ \b^{\cal P}_{\r\rho}\approx 26\ .\label{fa2}$

Using (\ref{p1}) in the approximation (\ref{P0}) violates unitarity.
We can observe from this that inelasticity effects are important in the determination
of the FSI phases \cite{Donoghue:1996hz}. As matter of fact, let us
 parametrize them as in Ref. \cite{Donoghue:1996hz} by one effective state,
 with no extra phases. Then
the $S$-matrix should be written as follows (neglecting a
small phase $\varphi=\,-0.01$ in $\sqrt{\cal P}$):
\be (B\to K^*\r)\hskip2cm S\approx\left(%
\begin{array}{cc}
  0.64 & 0.77 i \\
 0.77 i & 0.64 \\
\end{array}%
\right) \ ,\hskip1cm
 \sqrt{S}\approx\left(%
\begin{array}{cc}
  0.80& 0.62(1+ i) \\
 0.62(1+ i) & 0.80 \\
\end{array}%
\right) \ ;\ee

\be (B\to K^*\phi)\hskip2cm S\approx\left(%
\begin{array}{cc}
  0.77& 0.64\,i \\
 0.64\, i & 0.77 \\
\end{array}%
\right)\ , \hskip1.0cm \sqrt S\approx\left(%
\begin{array}{cc}
  0.88& 0.57(1+i) \\
  0.57(1+i) & 0.88 \\
\end{array}%
\right) \ ;\ee
\be (B\to \r\r)\hskip2cm  S\approx\left(%
\begin{array}{cc}
  0.58& 0.82\,i \\
 0.82\, i & 0.58 \\
\end{array}%
\right)\ , \hskip1.0cm \sqrt S\approx\left(%
\begin{array}{cc}
  0.76& 0.64(1+i) \\
  0.64(1+i) & 0.76 \\
\end{array}%
\right) \ .\ee
This shows that even neglecting the effect of
the non leading Regge trajectories, final state interactions
 due to inelastic effects parameterized by the Pomeron
 exchange can produce sizeable strong phases.
 This result agrees with the analogous findings of Refs.
 \cite{Nardulli:1997ht}  and \cite{Donoghue:1996hz}.

\subsection{Regge trajectories\label{reg}}
Let us now consider the contribution of the charmed Regge
trajectories. They are present in the $B\to K^*\phi$ amplitude, in
the
 $I=1/2$  amplitude  for the decay $B\to K^*\r$ and in $I=0$ amplitude
  for $B\to\r^+\r^-$.
 They do not contribute to the $B^+\to \r^+\r^0$ decay channel
because the final state has $I=2$ and the two charmed mesons can
only have either $I=0$ or $I=1$.

Including charmed Regge trajectories the  $S$ matrix can be
written for the generic  $B\to VV$ case as follows
\be
S={\cal
P}+{\cal D}+{\cal D}^*\, .
\label{regg}
\ee
Here ${\cal P}$ is the
Pomeron contribution discussed above; we note that ${\cal
P}_{ij}={\cal P}_{i}\delta_{ij}$. $ {\cal D}$ and ${\cal D}^*$ are
reggeized amplitudes corresponding to the $ {D}$ and ${D}^*$ Regge
trajectories. For $B\to K^*\phi$, $K^*\r$ decays they connect the
state $| K^* V\rangle$ to the other states $| D_s D\rangle$, $|
D_s D^*\rangle$, $| D^*_s D\rangle$, $| D^*_s D^*\rangle$. For
$B^0\to \r^+\r^-$ they connect the state $|\r^+\r^-\rangle$ to the
states $| D^{+} D^{-}\rangle$, $| D^{*+} D^{-}\rangle$, $| D^{+}
D^{*-}\rangle$, $| D^{*+} D^{*-}\rangle$.  Since all the Regge
contributions are exponentially suppressed one  can make the
approximation
\be
\sqrt{S}\approx \sqrt{\cal P}+\frac 1 2 {\cal
P}^{-1/2}\left({\cal D}+{\cal D}^*\right)\, .
\ee

We discuss explicitly the case $B\to K^* \phi$, the treatment of
the other cases  being similar. Neglecting other inelastic effects
we consider a $5\times 5$ $S_{ij}$ matrix, where $i=1,2,3,4,5$
represent respectively the states $| D_s D\rangle$, $| D_s
D^*\rangle$, $| D^*_s D\rangle$, $| D^*_s D^*\rangle$, $| K^* \phi
\rangle$. The kinematic factors are as follows:
$\ell_j=s(s-4m^2_D)\ ,j=1,\cdots,4$, $\ell_5=s^2$,
$\r_j=\sqrt{(s-4m^2_D)/s}\ ,j=1,\cdots,4$, $\r_5=1$. We work in
the approximation $m^2_D=m^2_{D_s}=m^2_{D^*}=m^2_{D^*_s}$ and
$m_{K^*}\,=\,m_\r=m_\phi\,=\,m_V\sim 0.9$ GeV with $s=m^2_B$. The integration limits
are for the elastic contribution: $t_-=0$ and $t_+=-s+4m_V^2$;  for
the inelastic contributions: $t_\mp=-s/2+m^2_V+m^2_D\pm s/2\sqrt{1-4m_V^2/s}
\sqrt{1-4m_D^2/s}$. Therefore
\be
A(B \rarr K^{*}\phi)_{\lambda}=
\sqrt{{\cal P}}A^{(5)}_{b\,,\lambda}\ +\
\sum_{k=1}^4\,\sum_{\lambda_i}\,\frac 1 2 {\cal
P}^{-1/2}\left({\cal D}\ +\ {\cal
D}^*\right)^{(k)}_{\lambda_i\,,\lambda} A_{b\,,\lambda_i}^{(k)}\, .
\ee
Let us consider the Regge amplitudes $R={\cal D}\ , {\cal D}^*$:
\be
R^{(k)}(s)=\frac{2i}{16\p}\sqrt[4]
{\frac{s-4m^2_D}{s}}\frac{1}{\sqrt{s(s-4m^2_D)}} \int_{t_+}^{t_-}
dt R^{(k)}(s,t)\, ,
\ee
with $ t_{\pm}$ given above. We assume the general parametrization
\starteq
R^{(k)}(s,t)= -\b^{R} {1 + (-)^{s_{R}}
e^{ -i\p\a_{R}(t)} \over 2}\G(l_{R} - \a_{R}(t))(\a ')^{1 - l_{R}}
(\a 's)^{\a _{R}(t)}
\label{R}
\fineq
as suggested in \cite{Irving:1977ea}. We notice the Regge poles
at $l_{R} - \a_{R}(t)=0,-1,-2,\cdots$; for
$R={\cal D}$ we have $s_R=\ell_R=0$ and,
 for $R= {\cal D}^*$,  $s_R=\ell_R=1$.
Near $t = m_{R}^{2}$, (\ref{R}) reduces to \starteq R \approx
\b^{R}{s^{s_{R}} \over ( t-m_{R}^{2} )} \label{Rapprox} \fineq
which allows to identify $\b^{R} $ as the product of two on-shell
coupling constants. Let us write \be \beta^R=\beta_{D_s^{(*)}
K^*}\beta_{D^{(*)} V}\ .\ee Using the effective lagrangian
approach \cite{Casalbuoni:1996pg} we can compute the residues by
identifying them with the coupling constants for $t\sim m^2_R$.
The numerical results are in
 Table \ref{table2}; they are obtained
 by the following values of the constants defined in \cite{Casalbuoni:1996pg}:
 $g_V=5.8$, $\l=0.56$
GeV$^{-1}$, $\b=0.9$, which represents an updated fit, see \cite{Isola:2003fh}
for details.

\vskip0.5cm
\begin{table}[ht]
\begin{center}
\begin {minipage}{6.5in}
\begin{tabular}{||c|c|c||c|c||}
\hline
Residue &  ${\cal D}$ & \rm{ Num. values} & ${\cal D}^*$&  \rm{ Num. values}\\
\hline
$\beta_{DV}^{0}$ & $ \beta\, g_V\,m_V/\sqrt 2$&  +2.84& 0 &0\\
\hline
$\beta_{DV}^{\pm}$\ &\ $\beta\, g_V\,m_D$ &$+9.76$&
$\sqrt 2 \lambda g_V\,m_D $ &$+8.59$\\
\hline
$\beta_{D^*V}^{0,0}$  &   $0$& $0
$ &$\displaystyle\frac{m_V\,g_V}{2\,m_D}(\beta-4\lambda m_D)
$&$-3.93$ \\
\hline
$\beta_{D^*V}^{\pm,\pm}$ &
$\sqrt{2}\,\lambda \,g_V\, m^2_D$ &  +16.1& $2\,\lambda\,g_V\, m_D
$ &$+12.1$\\
\hline
$\displaystyle\beta_{D^*V}^{0,\pm}$  &
$2\,\lambda\, g_V\,m^2_D  $ & +22.7 &
$\displaystyle \frac{g_V}{\sqrt 2}
\left(\beta-2\lambda m_D\right)
$ &$-4.90$ \\
\hline
$\beta_{D^*V}^{\pm,0}$  &
$ 2\,\lambda\, g_V\,m_V\,m_D  $& +9.35 &$0
$& 0\\
 \hline
$\beta_{D^*V}^{\pm,\mp}$  &  $\sqrt{2}\,\lambda \,g_V\, m^2_D$
&+16.1&$0  $&0 \\
  \hline
\end{tabular}
\end {minipage}
\end{center}
\caption{
Regge residues $\b^{\l_i,\l}$ of the poles  ${\cal D}^*$ and
${\cal D}$; $\l_i$ is the polarization of the charmed meson, $\l$
that of the light vector meson. The numerical results are obtained
with $g_V=5.8$, $\l=0.56$ GeV$^{-1}$, $\b=0.9$, see
\cite{Isola:2003fh}. $SU(3)_f$ nonet symmetry has been assumed for
the evaluation of the residues.\label{table2}}
\end{table}

It is important to stress that, differently form the Pomeron,
 which is mainly helicity conserving,
 Regge charmed trajectories have several helicity-flip residues.
  In particular they can change longitudinal
 into transverse polarizations, see e.g. $\displaystyle\beta_{D^*V}^{0,\pm}$. In the next section
 we will discuss these effects in the context of $B\to VV $ decays.

Next step is the calculation of the trajectories $\alpha_{D}(t)$
and $\alpha_{D^*}(t)$. To do this we use the model described in
the Appendix. For charmed mesons such as $D$ and $D^*$ (we do not
distinguish between $D$ and $D_s$)  the Regge trajectories are not
linear in the whole $t$ range. Due to the exponential suppression
of the high $|t|$ range, only  the small $t$ region is of interest
and here we can approximately write\be
\alpha_{D}=s_D+\alpha_0+\alpha^\prime t\,,\hskip1cm
\alpha_{D^*}=s_{D^*}+\alpha_0+\alpha^\prime t\ ,\ee with $s_D=0$,
$s_{D^*}=1$. From the results of the Appendix we have
$\alpha_0=-1.8$. As to the value of the slope $ \alpha^\prime$,
linearizing the trajectory as in the dashed-dotted line of Fig.
\ref{fig:2} (right side)
 one gets $\alpha^\prime=0.33$ GeV$^{-2}$.
This formula overestimates the mass of the $D,\,D^*$ system by 15\%.
Fitting these masses would give the value $\alpha^\prime=0.45$ GeV$^{-2}$;
from these results one can estimate a range 0.33-0.45 GeV$^{-2}$ for  $\alpha^\prime$.
In a conservative vein we double this range
to take into account other  theoretical uncertainties (arising from the model
adopted in the Appendix)
and use therefore
\be\alpha_0=-1.8\,,\hskip1cm
\alpha^\prime\,=\,(0.39\,\pm \,0.12) \ {\rm GeV} ^{-2}\ . \label{eq:27}\ee

\section{Numerical results and discussion\label{sec:2}}
In the present approach one can identify two kinds of theoretical
uncertainties, one related to the choice of the semileptonic $B\to
V$ form factors, the other to the parameters of the Regge
amplitudes. To start with we fix the form factors, by choosing the
numerical results  obtained by the Light Cone sum rules approach
\cite{Ball:2003rd}. The bare amplitudes can be found in Tables
\ref{table1bis} and \ref{table1ter}; for the Regge parameters we
use eq. (\ref{eq:27}), limiting the analysis to the central value
and the two extremes of the $\a^\prime$ range. The numerical
results are presented in Table \ref{Ball1}.

\begin{table}[ht]
\begin{center}
\begin{tabular}{|c|c|c|c|c|}
\hline
Process   & $\Gamma_L/\Gamma$&  $\Gamma_\perp/\Gamma$ &
$\Gamma_\parallel/\Gamma$  & B.R.\, $\times 10^6$  \\
\hline
$B\to K^*\phi $    &
$\begin{array}{c} 0.31\\0.73\\0.87\end{array}$  &
$\begin{array}{c} 0.1\\0.06\\0.05\end{array}$  &
$\begin{array}{c} 0.59\\0.20\\0.08\end{array}$  &
$\begin{array}{c} 35\\15\\12\end{array}$\\
\hline
$B\to K^{*+}\r^0$  &
$\begin{array}{c} 0.26\\0.72\\0.89\end{array}$  &
$\begin{array}{c} 0.12\\0.08\\0.06\end{array}$  &
$\begin{array}{c} 0.61\\0.20\\0.05\end{array}$  &
$\begin{array}{c} 19\\6.2\\4.9\end{array}$\\
\hline
$B\to K^{*0}\r^+$  &
$\begin{array}{c} 0.18\\0.61\\0.84\end{array}$  &
$\begin{array}{c} 0.11\\0.07\\0.05\end{array}$  &
$\begin{array}{c} 0.71\\0.32\\0.11\end{array}$  &
$\begin{array}{c} 35\\9.9\\7.2\end{array}$\\
\hline
$B^+\to\r^+\r^0 $ &
$0.95$  &
$0.03$  &
$0.03$  &
$14$\\
\hline
$B^0\to\r^+\r^- $ &
$\begin{array}{c} 0.86\\0.93\\0.94\end{array}$  &
$\begin{array}{c} 0.01\\0.02\\0.02\end{array}$  &
$\begin{array}{c} 0.13\\0.05\\0.04\end{array}$  &
$\begin{array}{c} 28\\26\\26\end{array}$\\
 \hline
\end{tabular}
\end{center}\caption{
Results for the various $B$ decay channels obtained with the form factors as in
\cite{Ball:2003rd}. For each decay channel the Regge slope of the charmed trajectories
is, from top to bottom, $\alpha^\prime$ = (0.27,0.39,0.51) GeV$^{-2}$.
There is no Regge contribution to $B^+\to\r^+\r^0 $ so that the only FSI effect here
is the elastic one. All the Branching Ratios are evaluated using $\tau_B\ =\ 1.67$ ps.\label{Ball1}}
\end{table}

We note that for $B\to K^*V$ decays the smallest value of
$\a^\prime$ corresponds to the largest contribution from the Regge
trajectories. The Table \ref{Ball1} shows that the
 main effect of the Regge poles for the channels with a $K^*$ in the final state
 is the reduction
of the longitudinal polarization fraction and the increase of the
other fractions. This is due to the effect mentioned in subsection
\ref{reg}, i.e. the existence of helicity flip residues changing
longitudinal into transverse polarizations. On the other hand the
transverse polarization fraction $\G_\perp/\G$ cannot increase
beyond 10\%-12\%.

\begin{figure}[ht]
\centerline{
\begin{tabular}{cc}
\epsfig{file=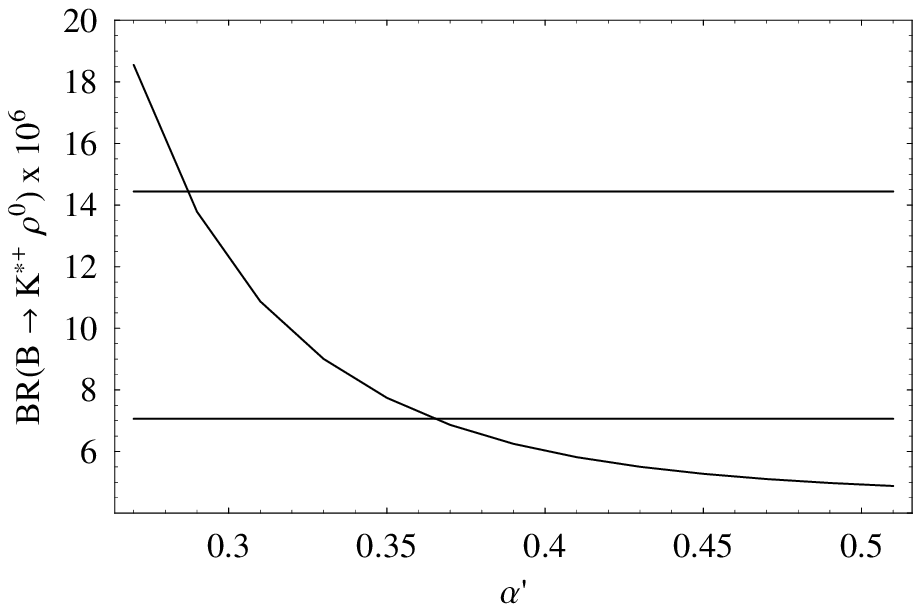,width=8truecm } &
\epsfig{file=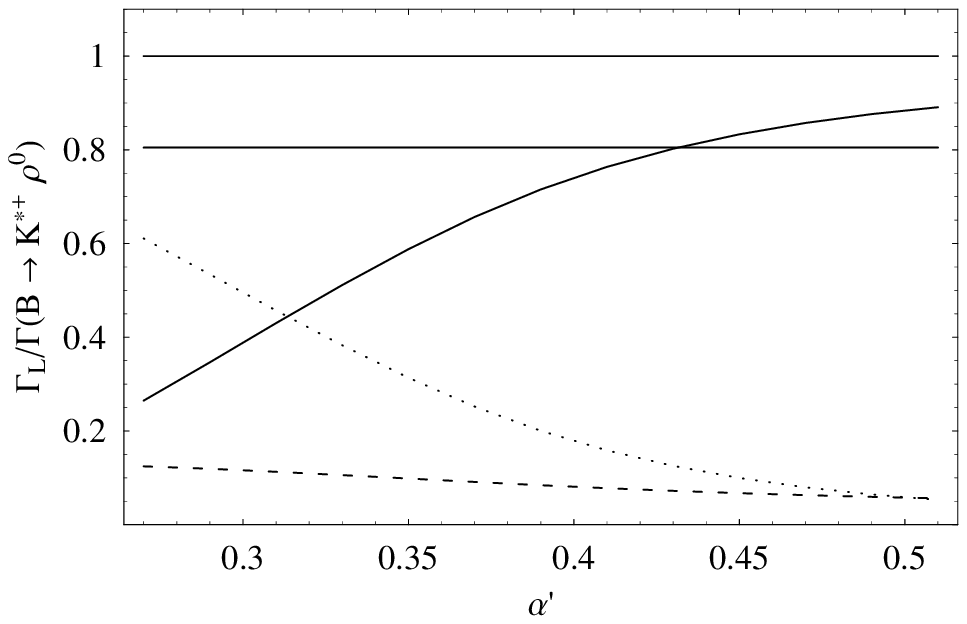,width=8truecm  }
\end{tabular}}
\caption{Comparison between theoretical predictions based on the
form factors of Ref. \cite{Ball:2003rd} and the experimental data
for the channel $B^+\to K^{+*}\rho^0$.  On the left: The
$B.R.\times 10^6$ as a function of the parameter $\alpha^\prime$
giving the slope of the charmed Regge trajectories. Continuous
straight lines give the experimental interval (see Table
\ref{table0}). On the right: Continuous lines give the fraction
$\G_L/\G$, the dashed line is the fraction $\G_\perp/\G$ and the
dotted line the fraction $\G_\parallel/\G$. Units of
$\alpha^\prime$ are GeV$^{-2}$. We use $\tau_B\ =\ 1.67$ ps. \label{figkstar}}
\end{figure}

A more detailed numerical analysis for the channel $B\to
K^{*+}\r^0$ can be found in Fig. \ref{figkstar} that shows the
variation of the branching ratio and the polarization fractions
versus the slope $\a^\prime$. A detailed analysis of the $K^*\phi$
channel will be presented  below. Finally we observe that, as
expected, the $\r\r$ channels get negligible or vanishing
contribution from the Regge poles. For the $\r^+\r^-$ channel this
is due to the absence of the Cabibbo enhancement; for the
$\r^+\r^0$ decay mode there is no Regge contribution because the
final state  is a pure $I=2$ state. If the explanation presented
here is valid, the large fraction $\G_L/\G$ for $B\to\r\r$ (see
Table \ref{table0}) is a consequence of the negligible role of the
charmed Regge poles.

\begin{figure}[ht]
\centerline{ \epsfig{file=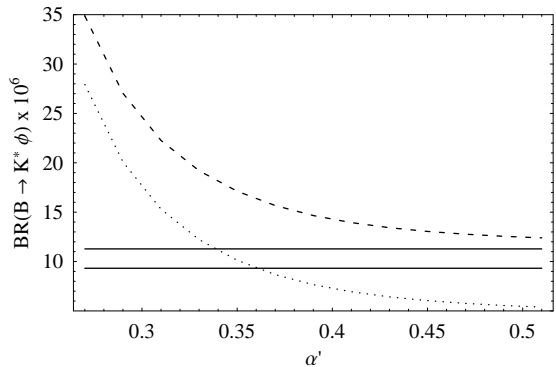,width=8truecm }
} \caption{ $BR(B\to K^*\phi)\times 10^6$ as a function of the
parameter $\alpha^\prime$ giving the slope of the charmed Regge
trajectories. Continuous straight lines give the experimental
interval (see Table \ref{table0}). Dashed lines are theoretical
predictions based on the form factors of ref.\cite{Ball:2003rd};
dotted lines are based on the form factors of ref.
\cite{Colangelo:1996jv}. Units of $\alpha^\prime$ are GeV$^{-2}$.
\label{col1}}
\end{figure}

\begin{figure}[ht]
\centerline{ \epsfig{file=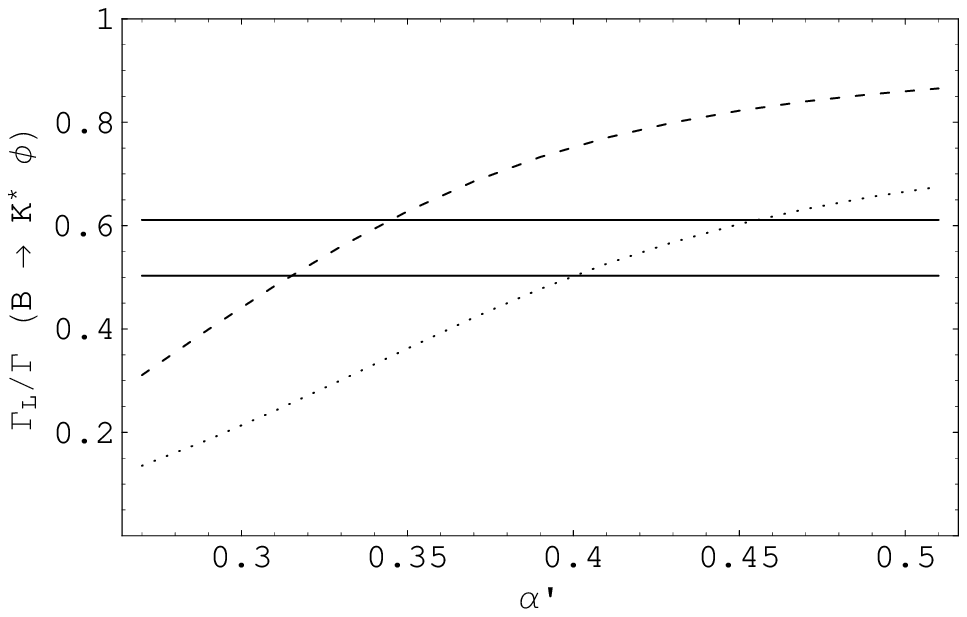,width=8truecm }
} \caption{ $\G_L/\G (B\to K^*\phi)$ as a function of the
parameter $\alpha^\prime$ giving the slope of the charmed Regge
trajectories. Continuous straight lines give the experimental
interval (see Table \ref{table0}). Dashed lines are theoretical
predictions based on the form factors of ref.\cite{Ball:2003rd};
dotted lines are based on the form factors of ref.
\cite{Colangelo:1996jv}. Units of $\alpha^\prime$ are
GeV$^{-2}$. We use $\tau_B\ =\ 1.67$ ps. \label{col2}}
\end{figure}

Let us comment on the discrepancies between the theoretical
results of Table \ref{Ball1}, Fig. \ref{figkstar} and experimental
data. They might point to new physics, but a simpler possibility
is an interplay between final state interactions and form factors.
An example is offered by the channel $K^*\phi$. Here to get a
small longitudinal polarization fraction one would prefer a small
value of $\a^\prime$. Small values of $\a^\prime$, however,
produce  values for the branching ratio much higher than the data.
As discussed in the Introduction, the dominance of $\G_L/\G$ is a
consequence of the chiral structure of the currents in the
standard model and the $m_b\to\infty$ limit. The only way to
reduce it without invoking new physics is to show the existence of
soft effects in the standard model. Our analysis shows that these
effects can be generated by rescattering effects parametrized  by
the Regge model.  On the other hand the high value of the
branching ratio might be due to the choice  of the form factors of
Ref. \cite{Ball:2003rd}. To deal with this problem we consider a
different set of form factors. Since in $B\to VV$ decays only
small $q^2$ are involved, the Lattice QCD determinations are
unreliable because of the large extrapolations they would need.
The best alternative to Light Cone sum rules are the traditional
QCD sum rules. In \cite{Colangelo:1996jv} such a calculation is
performed; we have reported these predictions in Table
\ref{table1bis}. The authors in \cite{Colangelo:1996jv} only
consider the $b\to s$ transition, and therefore their results can
be only applied to the channel $B\to \phi K^*$ (in \cite{Ball:1997rj}
also the transition $b\to u$ was considered, but with
a set of parameters different from \cite{Colangelo:1996jv}, which
renders the comparison difficult). In figures \ref{col1} and \ref{col2}
we report the branching ratio and  the fraction
$\G_L/\G$ as functions of the slope $\a^\prime$ of the charmed Regge
trajectories for the parametrizations \cite{Colangelo:1996jv} and \cite{Ball:2003rd}.
As to the other  polarization fractions , $\G_\parallel/\G$ and
$\G_\perp/\G$ for the form factors of Ref. \cite{Ball:2003rd} can
be found in Table \ref{Ball1}. The analogous figures for the form
factors of  \cite{Colangelo:1996jv} are
$\G_\perp/\G=(0.12,\,0.12,\,0.12)$ and
$\G_\parallel/\G=(0.74,\,0.40,\,0.20)$ for
$\a^\prime=(0.27,\,0.39,\,0.51)$ GeV$^{-2}$.

Fig. \ref{col1} shows that much smaller values of the branching
ratio are  obtained by the QCD sum rules form factors  of Ref.
\cite{Colangelo:1996jv} (dotted lines) for $\a^\prime$ around its
central value; Fig \ref{col2} shows that for the same values of
$\a^\prime$ satisfactory results for  $\G_L/\G$ are obtained.
Finally $\G_\perp/\G$ gets a value smaller, but still compatible
within 2$\s$, with the BaBar result (\ref{babarr}).

These results show that data on the polarization fractions
 for the $B\to K^*\phi$ decay
channel might be explained as the effect of two converging
factors: an appropriate set of form factors
\cite{Colangelo:1996jv} and the presence of final state
interactions computed by the Regge approach.  It remains to be
seen however how QCD sum rules and Regge phenomenology would work
for the other channels $B\to K^*\r$ and $B\to\r\r$, an issue we
plan to deal with in the future.

\section{Appendix: a model for Regge trajectories \label{app:0}}

 For non charmed Regge poles,
$\a_{R}(t)$ is phenomenologically
given by
\starteq
 \a_{R}(t) = s_R + \a '(t -m^2_R)
\label{alphaR}
\fineq with the
 universal slope: $\a '= 0.93 \, \rm GeV^{-2}$.
 This behavior can be
  reproduced by a potential model, which  gives the Regge formula
  the meaning of an exchange of infinitely
  many particles. To compute their  spectrum one solves a Schroedinger equation
  and one shows that the slope is related to the string tension
  responsible of the confining part of the potential.

To be more specific we consider a bound state comprising a quark and an antiquark.
For the bound state problem we consider the Salpeter
equation (Schr\"odinger equation with relativistic kinematics):
\be
\left(\sqrt{{\bf p_1}^2+m_1^2}+\sqrt{{\bf p_2}^2+m_2^2}+V(r)\right)\psi({\bf r })
= E\psi({\bf r})
\ee
where $m_j$ are quark masses, $V(r)$ is the potential
energy  and $r=|{\bf r_1}-{\bf r_2}|$.
In the meson rest frame $E$ is the meson mass $M$ and
${\bf p_1}^2={\bf p_2}^2=-\hbar^2\nabla^2$.
The bound state equation can be solved for arbitrarily large
quantum numbers $n$ and $\ell$ (radial and orbital quantum numbers)
by the WKB method \cite{Cea:1982rg}. We assume the potential
\be
V(r)=V_0+\mu^2 r
\label{vo}
\ee
which is of course a simplification, but however able to show
the linearity of Regge trajectories; $\mu$ is the string tension and
$V_o$ is a constant negative term mimicking a repulsive core.

\begin{figure}[ht]
\centerline{
\epsfxsize=6cm\epsfbox{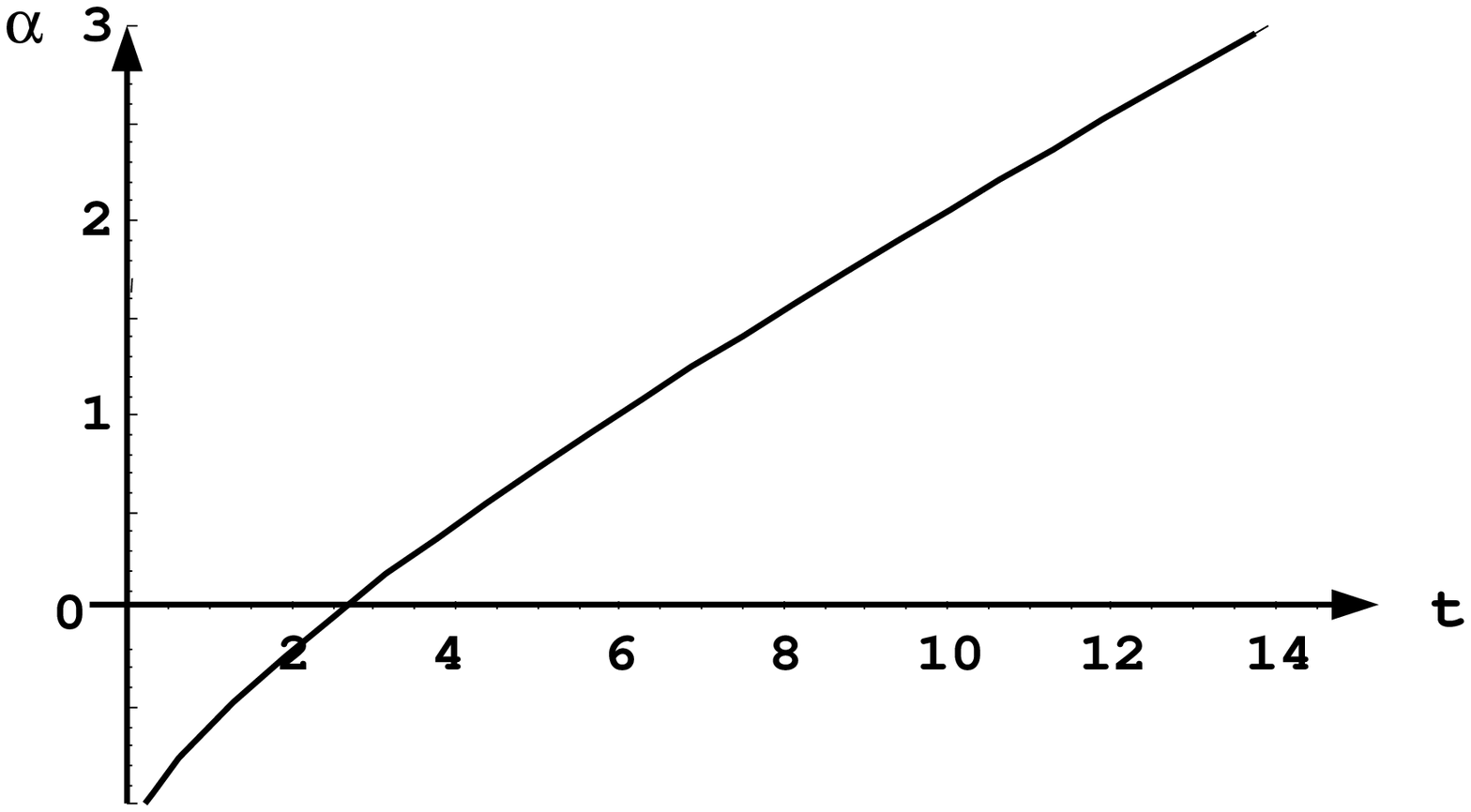}\hskip1cm\epsfxsize=6.5cm\epsfbox{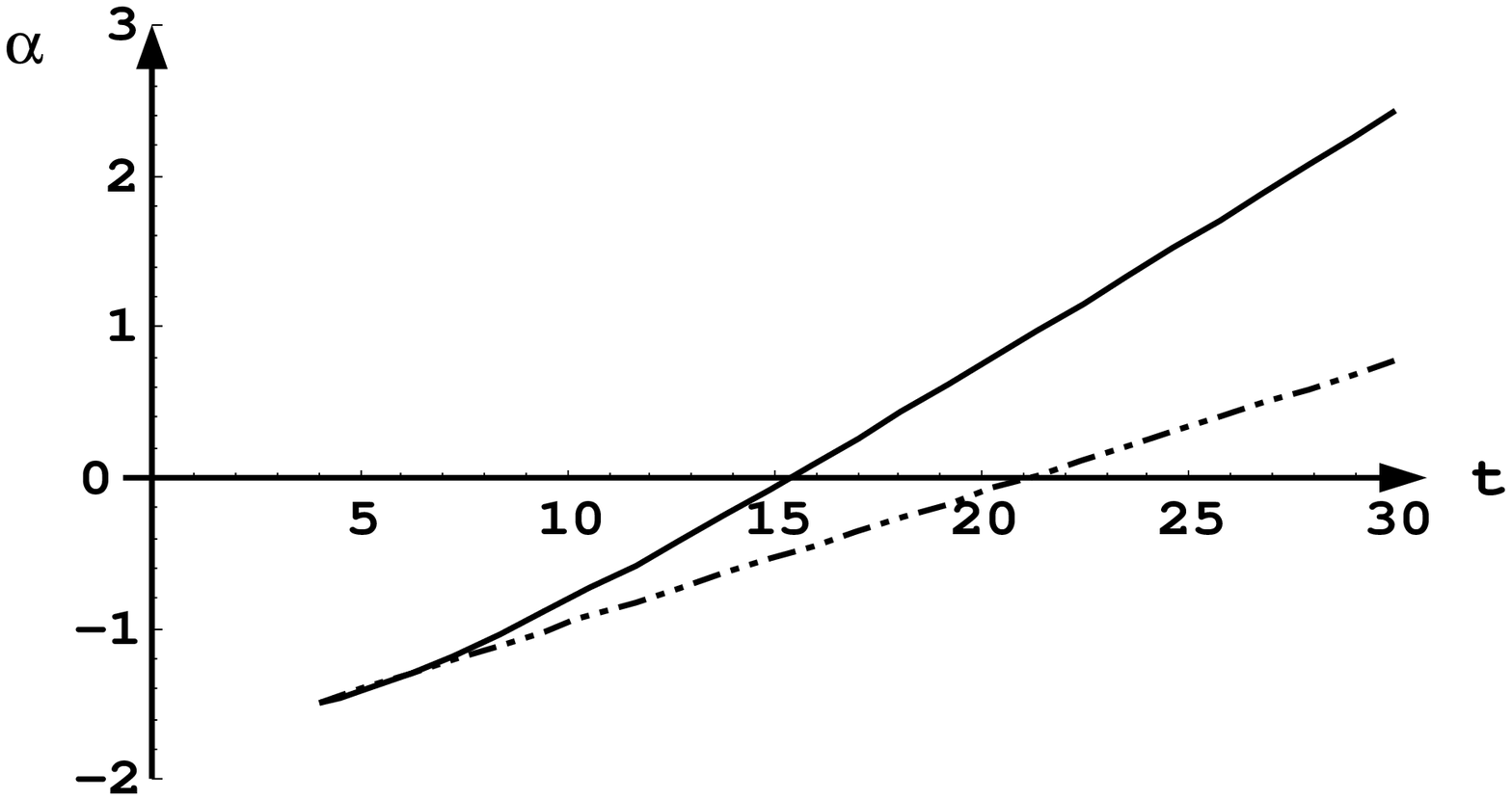} }
\caption {{ On the left: Chew-Frautschi plot for a $q\bar q$ meson of mass $M$
 made up by light quarks with
masses of 0.577 in units of string tension ($\sim 300$ MeV). On the horizontal
axes $t=M^2$ in units of string tension, on the
the vertical axis the Regge trajectory
$\alpha(t)$ approximately given by
$\alpha(t)=s_R+\alpha^\prime (t-m^2_R)$ with $s_R=0$, see text. The meson masses
correspond to successive values of $\alpha=\ell=0\,,1\,,\cdots$.
On the right: Chew-Frautschi plot for a $q\bar Q$ meson of mass $M$
made up by a charm quark and a light quark ($m_c=1.7 $ GeV, $m_q=0.3 $ GeV).
On the horizontal axes $t=M^2$ in units of string tension, on the
the vertical axis the Regge trajectory. Solid line gives the
WKB solution of the bound state equation; dashed-dotted line a
linear approximation in the small $t$ region.
\label{fig:2} }}
\end{figure}

The WKB formula for this potential gives the result
\cite{Cea:1982rg} (in units of string tension)
\be \displaystyle
\int_0^\sigma dr\ \sqrt{\frac{(M-V(r))^2}{4}-\frac{m_1^2+m_2^2}{2}
+\frac{(m_1^2-m_2^2)^2}{4(M-V(r))^2}}\ =\ \pi\left( n+\frac \ell 2
+ \frac 3 4\right)\hskip.2cm\
\ee
where $\sigma=M-V_0-m_1-m_2$. For
light quarks we take for the constituent quark masses the values
$m_1=m_2=300$ MeV. In units of string tension we get, for
$V_0=-1.923$ and $n=0$ the Chew-Frautschi plot of Fig.
\ref{fig:2} (left side) showing the almost linear Regge
trajectory. It is given by Eq. (\ref{alphaR}) with $s_R=0$ since
no spin term has been added to the potential in (\ref{vo}). For
$\alpha=\ell$ we find successive squared meson masses. The Regge
trajectory has slope $\alpha^\prime=0.25/\mu^2$. For $\mu=0.52$
GeV we get a phenomenologically acceptable value of
$\alpha^\prime\sim 0.9$ GeV$^{-2}$. The mass of low-lying meson
with $\ell=0$ made up by up/down quarks turns out to be $\sim 800$
MeV, an acceptable value as well.

We can now repeat the exercise for the $D, D^*$ mesons. We use again  $m_1=300 $ MeV and
put for the charm quark $m_2=m_c=1.7 $ GeV. The result is reported on the right in
Fig. \ref{fig:2}. One can observe some deviations from linearity.
The minimum value is given by kinematics as
$t=4$ in units of the string tension.  Solid line gives the
WKB solution of the bound state equation; for small values we can still approximate
$\alpha(t)$ by a straight line (dashed-dotted line in the figure),
with $\alpha(t)=\alpha_0+\alpha^\prime t$
and $\alpha_0=-1.8$, $\alpha^\prime= 0.09/\mu^2=0.33 $ GeV$^{-2}$.


\end{document}